\documentclass{jmlr}

\usepackage[utf8]{inputenc}
\usepackage[T1]{fontenc}
\usepackage{lmodern}
\usepackage{booktabs}
\usepackage{array}
\usepackage{microtype}

\makeatletter
\long\def\@makecaption#1#2{%
   \vskip 6pt
   \small
   \setbox\@tempboxa\hbox{#1: #2}%
   \ifdim \wd\@tempboxa >\hsize
       \begin{list}{#1:}{%
       \settowidth{\labelwidth}{#1:}
       \setlength{\leftmargin}{\labelwidth}
       \addtolength{\leftmargin}{\labelsep}
        }\item #2 \end{list}\par
     \else
       \hbox to\hsize{\hfil\box\@tempboxa\hfil}
   \fi}
\renewcommand*{\@jmlrenddoc}{%
  \FloatBarrier
  \phantomsection
  \protected@edef\@currentlabelname{end of \@shorttitle}%
  \label{jmlrend}\null
  \global\let\@reprint\@empty}
\renewcommand{\ps@jmlrtps}{%
  \let\@mkboth\@gobbletwo
  \def\@oddhead{}\let\@evenhead\@oddhead
  \def\@oddfoot{\@titlefoot}\let\@evenfoot\@oddfoot}
\makeatother

\newcommand{\widepaperfig}[1]{\makebox[\linewidth][c]{\includegraphics[width=1.04\linewidth]{#1}}}

\jmlryear{2026}

\title[Cost-Aware Security Agent Evaluation]{Beyond Success Rate: Cost-Aware Evaluation of Offensive and Defensive Security Agents}

\author{\Name{Paul Kassianik}
\AND
\Name{Blaine Nelson}
\AND
\Name{Yaron Singer}}

\hypersetup{
  pdftitle={Beyond Success Rate: Cost-Aware Evaluation of Offensive and Defensive Security Agents},
  pdfsubject={Cybersecurity benchmark evaluation},
  pdfauthor={Paul Kassianik, Blaine Nelson, and Yaron Singer},
  pdfkeywords={cybersecurity benchmarks, AI agents, Cybench, BOTS, SOC, cost accounting}
}

\newcommand{\BOTSvOne}{BOTS~v1}
\newcommand{\Inspect}{Inspect}

\begin{document}
\maketitle
\begingroup
\renewcommand{\thefootnote}{}
\renewcommand{\footnoteseptext}{}
\footnotetext{Correspondence to:
\href{mailto:pkassianik@gmail.com}{\texttt{pkassianik@gmail.com}};
Twitter: \href{https://twitter.com/kass_paul}{\texttt{@kass\_paul}}.}
\endgroup

\begin{abstract}
Security-agent evaluations commonly measure peak offensive capability under generous inference budgets, emphasizing vulnerability discovery, exploit development, penetration testing, and CTF completion.
Such measurements are useful but incomplete: in operational security, every reasoning step, tool call, telemetry query, and enrichment request consumes budget.
We evaluate language-model security agents through this cost-success lens on offensive Cybench challenges and defensive Splunk BOTS v1 investigation challenges.
Instead of reporting only best-case success, we compare models at fixed cost levels and decompose performance by inference spend and tool spend.
Our results show distinct scaling regimes for red- and blue-team tasks.
Offensive CTF performance improves with additional test-time compute, and scaled open-weight models can approach frontier proprietary systems while remaining cost-competitive.
Defensive SOC investigation does not scale in the same way: success depends more heavily on disciplined tool use, telemetry navigation, and selective enrichment than on raw reasoning budget alone.
We argue that security-agent benchmarks should measure economic efficiency and operational fit alongside task success.
Cost-aware, SOC-native evaluations provide a clearer picture of which models are practically useful today and where defensive agents still need to improve. We present an interactive website with our results as \href{https://evals.frontier.security}{https://evals.frontier.security}.
\end{abstract}


\section{Introduction}
\label{sec:introduction}

\begin{figure}[t!]
\centering
\widepaperfig{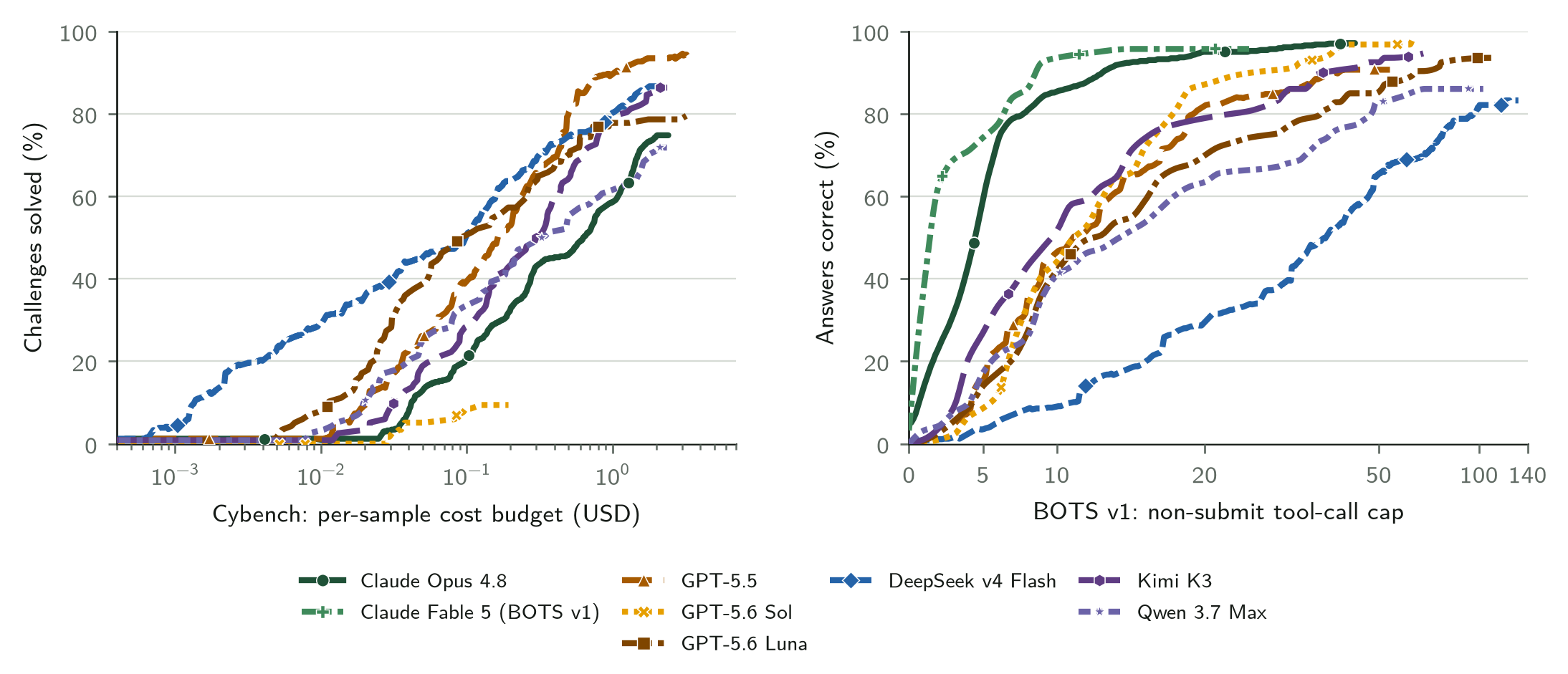}
\caption{Resource scaling for various models on Cybench and BOTS v1. \textbf{Left:} Cybench success rate versus a per-sample cost budget cap: a challenge counts as solved at budget $x$ only if its model-token spend was at most $x$. \textbf{Right:} BOTS v1 binary answer accuracy versus a per-sample cap on tool calls, shown on a symmetric-log axis. Claude Fable 5 appears only BOTS v1 due to refusals on Cybench. These GPT-5.6, Fable, and Kimi K3 runs use \texttt{high} reasoning effort. Offensive success climbs steadily with spend, while defensively the strongest models are near-maximal within ${\sim}20$ tool calls. Notably, DeepSeek v4 Flash does over 100 tool-calls to plateau lower --- defense tracks tool discipline, not volume.}
\label{fig:main-results-chart}
\end{figure}

Security-agent evaluations often ask how far a model can go when it is given a generous inference budget and an offensive task: solve a CTF, find a vulnerability, write an exploit, or automate a penetration-testing step.
Recent Claude Mythos Preview evaluations exemplify this regime, combining CTF suites, multi-step cyber ranges, and vulnerability-discovery/exploit-development evaluations with million-token budgets and attacker-directed execution access \citep{aisi2026mythos,anthropic2026mythos}.
Those measurements are useful because they track capabilities that attackers can use.
However, they are incomplete because operational security is not a peak-capability exercise.
Measuring capability at a cost limit is much more indicative of how these measurements can be valuable in production systems.
A security operations center (SOC) needs agents that can navigate telemetry, choose searches, decide when enrichment is worth paying for, and return analyst-grade answers under a budget.

We therefore make cost the common axis for comparing offensive and defensive security agents.
A fixed budget turns success rate into an operational question: how much capability does a model buy per dollar of inference and tooling?
This framing also separates two costs that are often collapsed together.
For offensive CTFs, most marginal spend is test-time reasoning and command execution.
For SOC investigation, the model also spends analyst-facing resources: Splunk queries, event inspection, web search, WHOIS history, and other enrichment calls.

We evaluate this cost-success tradeoff on offensive Cybench challenges \citep{zhang2024cybench} and defensive Splunk Boss of the SOC (BOTS) v1 investigations \citep{splunk_botsv1} in a common agent-evaluation harness.
Cybench provides sandboxed CTF-style tasks with shell and Python access.
BOTS v1 provides public Splunk telemetry and official competition questions with point values and hint penalties.
Rather than presenting a single leaderboard, we characterize operating points: model, task family, cost limit, tools, and scoring rule.

We find distinct scaling regimes.
On Cybench, retrospective budget caps show clear headroom for Claude Opus 4.8 and DeepSeek v4 Flash, while GPT-5.5 already solves most successful challenges below the lower cap.
On BOTS v1, extra budget is less predictive: the direct DeepSeek v4 Flash cap increase produces little gain, and high tool volume often accompanies lower scores.
The defensive signal is therefore not raw reasoning budget alone; it is disciplined tool use, telemetry navigation, and selective enrichment.

\noindent\begin{minipage}{\linewidth}
\hspace*{\parindent}Our contributions are as follows:
\begin{itemize}
  \item A cost-success comparison of shared Cybench and BOTS v1 model operating points using a common harness;
  \item A paired budget-headroom analysis showing where additional per-sample spend changes observed scores;
  \item Model-token plus priced-tool cost accounting for SOC investigation, including cost per 1,000 BOTS points;
  \item No-tools controls showing substantial model-dependent direct answer recovery on public BOTS v1 and therefore the need for decontamination checks before interpreting absolute scores.
\end{itemize}
\end{minipage}

\section{Related Work}
\label{sec:related-work}

Cybersecurity model evaluation spans knowledge tests, offensive agent tasks, defensive investigations, and benchmark-integrity work.
Broad suites such as CyberSecEval 2 measure security risks and capabilities including prompt injection, code-interpreter abuse, false refusals, and exploit-generation tasks \citep{bhatt2024cyberseceval2}; CyberSecEval 3 extends that line toward automated social engineering and autonomous offensive operations \citep{wan2024cyberseceval3}.
Cybersecurity-knowledge benchmarks such as SecBench scale up multi-domain question answering and short-answer grading \citep{jing2025secbench}.
These resources are useful for measuring domain knowledge and misuse risk, but they usually do not require a full investigation loop with changing evidence, paid tools, and explicit budget tradeoffs.

Offensive agent benchmarks add interaction and execution.
InterCode standardizes code tasks with execution feedback \citep{yang2023intercode}; Language Agents as Hackers introduced InterCode-CTF for capture-the-flag tasks \citep{yang2023languagehackers}; NYU CTF Bench broadens open CTF coverage with automated tool use \citep{shao2024nyuctfbench}; and Cybench uses professional CTF tasks with sandboxed command execution \citep{zhang2024cybench}.
CVE-Bench moves toward real-world web-application exploitation rather than abstract CTFs \citep{zhu2025cvebench}, while PentestGPT evaluates LLM support across penetration-testing subtasks \citep{deng2024pentestgpt}.
Recent InterCode-CTF results show that prompting, tool use, and multiple attempts can substantially change observed offensive performance \citep{turtayev2024hackingctfs}; EnIGMA further shows that interactive debugger and connection tools can substantially assist vulnerability-finding agents \citep{abramovich2025enigma}.
BountyBench broadens the unit of evaluation to detect, exploit, and patch tasks on real-world codebases with dollar-valued bug bounties \citep{zhang2025bountybench}.
We follow that agentic line, but evaluate cost-success operating points rather than peak success rate alone.

SOC evaluation draws on a different body of practice.
Incident-response guidance and cybersecurity workforce taxonomies emphasize detection, analysis, evidence handling, response coordination, and operational judgment \citep{nist_sp800_61r3,nist_nice_800_181r1}.
Splunk's BOTS datasets provide public security telemetry, questions, answers, and scoring artifacts for blue-team investigation exercises \citep{splunk_botsv1,splunk_botsv2,splunk_botsv3}.
An analyst must discover log schemas, write and revise SPL, inspect representative events, correlate artifacts, and decide whether external enrichment will change the case.
A flag-only score or a raw tool-call count cannot distinguish those behaviors.

Recent defensive-agent benchmarks are beginning to target this gap.
SIABench frames security incident analysis as alert triage plus deeper multi-step workflows across malware, phishing, forensic, and log-analysis scenarios \citep{majumdar2026siabench}.
ExCyTIn-Bench derives cyber-threat-investigation questions from graph-linked security logs \citep{wu2025excytin}; CyberTeam models blue-team threat hunting as embodied function-calling workflows \citep{liu2025cyberteam}; and Cyber Defense Benchmark evaluates open-ended threat hunting over raw Windows event logs \citep{kumar2026cyberdefensebench}.
Our BOTS v1 study complements this emerging line by preserving official BOTS scoring while adding fixed-budget comparisons, model-token cost, priced external-tool spend, and tool-efficiency views.

Public benchmarks also require leakage controls.
Benchmark-contamination studies show that public test sets can overlap with model-training data or be partially recoverable from model behavior, inflating apparent capability and complicating fair comparisons \citep{deng2024datacontamination,wang2024benchmarkleakage,li2024contaminationsurvey}.
Practical mitigation work recommends protected test releases, training-exclusion controls, private or retroactive holdouts, and care with API-mediated leakage \citep{jacovi2023testdata,balloccu2024leak,haimes2024benchmarkinflation}.
Because BOTS v1 is public and old, Section~\ref{sec:bots-contamination} treats no-tools and no-context probes as contamination controls rather than as primary capability results.

\section{Evaluation Design}
\label{sec:evaluation-design}

We use the Inspect evaluation framework \citep{inspect_ai_2024} to organize the experiments.
We use a ReAct-style agent with auto-compaction \citep{yao2023react} for all benchmark runs.
We auto-compacted when the agent context reached 90\% of the model context window.
Auto-compaction used Inspect AI's automatic compaction strategy, which attempts provider-native compaction and falls back to summary-based compaction \citep{inspect_ai_compaction_source}.
Between offensive and defensive tasks, we only change the suite of available tools.
We rely on \Inspect{} to keep track of auxiliary metrics like cost, token consumption, and tool calling statistics.
We set budgets on the cost as the main resource control on agent runs.
The agents are unaware of cost or token budgets but are aware of limits on certain expensive tool calls.

This design allows us to ascertain the model's success rate at various cost and token consumption caps.
We can derive practical metrics such as cost per solved question.
By varying the cost between 0 and the run limit, we can track how success rate scales with the cost and token budget increases.

For models provided by OpenAI and Anthropic, we use their respective APIs.
\Inspect{} also relies on frontier providers for autocompaction, prompt caching, and cost tracking metrics.
For other models, we use OpenRouter to select one primary provider and two fallback providers.
We use average pricing across the three providers to calculate the cost of runs.
Since OpenRouter delegates prompt caching to providers, having one primary provider enables a maximally efficient caching scheme without sacrificing operational stability.
Operationally, most OpenRouter model calls record some prompt-cache reads, though cache-miss rates vary by run.

\subsection{Offensive benchmark: Cybench}
\label{sec:cybench-method}

Cybench is an offensive CTF benchmark with 39 challenges in the full set, drawn from four cybersecurity competitions (including HackTheBox, Glacier, SEKAI, and HKCert) and spanning domains such as cryptography, web exploitation, reverse engineering, forensics, and exploitation \citep{inspect_evals_cybench}.
Each sample requires an agent to solve a sandboxed offensive security challenge (web/app exploitation, binary/pwn, reversing, crypto, web/forensic workflows, and other hands-on tasks) and submit a hidden challenge flag token (typically brace-delimited strings such as \texttt{HTB\{\dots\}}, \texttt{gctf\{\dots\}}, or \texttt{SEKAI\{\dots\}}).
Submitted answers are judged with a case-insensitive substring match against the reference flag; each challenge allows up to 3 submission attempts and we score by averaging binary correctness across three independent epochs (per-sample mean correctness, then averaged across all challenges).

We use the Cybench hard variant from Inspect Evals \citep{inspect_evals_cybench} to perform the evaluations.
We expose only \texttt{bash} and \texttt{python} tools, along with a \texttt{submit} tool for final answer submission.
All tools are considered to be zero-cost.
We limited the number of submission attempts to 3.
\subsection{Defensive benchmark: BOTS v1}
\label{sec:bots-method}

\BOTSvOne{} is a public Splunk incident-investigation dataset and CTF platform for information security professionals, researchers, students, and enthusiasts \citep{splunk_botsv1}.
We did not identify a public Inspect implementation of \BOTSvOne{}, so we implemented one.
Unlike a flat set of independent binary questions, the official BOTS format is a sequential, point-based investigation in which later questions can depend on earlier findings.
\BOTSvOne{} contains a warm-up plus 31 scored questions covering the Po1s0n1vy web-defacement investigation and Cerber ransomware at Wayne Enterprises.
Each scored question has official base points and, where applicable, ordered hints with point costs.
After an incorrect submission the next hint is revealed; a later correct answer earns base points minus consumed hint costs, while an unanswered or still-incorrect question earns zero.

We exclude the official warm-up question, leaving 31 scored questions worth 10,300 official competition points.
Agents receive Splunk discovery, search, and event-inspection tools over Splunk Search Processing Language \citep{splunk_search_reference}, plus limited VirusTotal, WHOIS, DNS, Brave Search, bash, and Python tools \citep{virustotal_api,whoisxmlapi_whois_history,brave_search_api,rfc3912,rfc1035}.
We make Splunk telemetry the primary evidence source in the prompt and ask agents to use external enrichment selectively.
To preserve cost efficiency, we limit VirusTotal and WhoisXMLAPI calls to 3 while limiting web search to 5 calls.
We implement strict keyword filtering to prevent competition writeups and solution leakage.

We use \texttt{bots\_points} as the primary metric: a correct answer earns official points minus hint penalties; an incorrect answer earns zero.
Hints are revealed with every unsuccessful answer.
We report binary \texttt{includes} accuracy as a secondary view.
We use three epochs and mean reduction, so point totals are reduced back to the single-epoch 10,300-point scale.
We provide prerequisite case context from earlier official questions for 23 of the 31 scored questions, matching the sequential investigation structure but making the task closer to follow-up analysis than cold-start incident reconstruction.

\subsection{Refusal accounting}
\label{sec:refusal-accounting}

We classify refusals at the sample-epoch level using both agent text and provider metadata.
During evaluation, an assistant turn is treated as a refusal when it contains nonblank text, makes no tool call, and matches a case-insensitive set of apology, inability, or policy phrases after Unicode normalization, or when the provider returns \texttt{content\_filter}.
Both benchmark agents use a refusal limit of one: the first detection that reaches the agent continuation hook aborts that sample, which is then scored on its current output.
This text detector is a heuristic rather than a provider policy label.

For reporting, let $N$ be the number of loaded scored sample-epochs and $R$ the number classified as refusals.
Cybench counts only failed sample-epochs whose final assistant turn matches the text detector or whose model-event log contains a content filter; it reports refusal rate $100R/N$, or $R/3$ challenge-equivalents for three-epoch runs.
Raw Cybench pass@1 is unchanged, so refused epochs remain failures.
BOTS v1 instead flags any matching assistant turn or content-filter event in the transcript and reports $R/N$ as secondary context without changing official points or binary accuracy.
A BOTS v1 refusal event can therefore coexist with points if a later or retried attempt answers correctly.

\subsection{Cost accounting}
\label{sec:cost-accounting}

Table~\ref{tab:cost-pricing} summarizes the pricing assumptions used in the reported dollar totals.

\begin{table}[t]
\centering
\caption{Cost-accounting assumptions for model and tool spend.}
\label{tab:cost-pricing}
\small
\begin{tabular}{@{}>{\raggedright\arraybackslash}p{0.13\linewidth}>
                  {\raggedright\arraybackslash}p{0.30\linewidth}>
                  {\raggedright\arraybackslash}p{0.33\linewidth}>
                  {\raggedright\arraybackslash}p{0.14\linewidth}@{}}
\toprule
Availability & Cost item & Unit price in reports & Per-question limit \\
\midrule
Shared & Model inference & Provider ledger or registered per-token rate sheet & Cost cap only \\
 & \texttt{bash}, \texttt{python} & \$0 marginal cost & None \\
\midrule
BOTS v1 only & Splunk discovery & \$0 marginal cost & None \\
 & Splunk search  & \$0 marginal cost & None \\
 & Splunk event drill-down & \$0 marginal cost & None \\
\cmidrule(l){2-4}
 & Brave Search & \$0.005 per request & 5 calls \\
 & WhoisXMLAPI history preview & \$0.0258 per request & 3 total calls \\
 & WhoisXMLAPI history purchase & \$1.29 per request & 3 total calls \\
 & VirusTotal public API & \$0 marginal cost & 3 calls \\
 & DNS, live WHOIS/RDAP & \$0 marginal cost & None \\
\bottomrule
\end{tabular}
\end{table}

In order to manage overall costs, we cap the cost for each sample run.
We set the cap via \Inspect's budget option -- this limits the runtime of a benchmark sample evaluation to a certain cost and aborts the agent if it overruns the budget.
We treat such examples as failed.
Most successful models finish well below the cost budget cap.

We decompose cost into inference spend and priced-tool spend.
For Cybench, reported cost is model-token cost recorded by the evaluation harness or reconstructed from token counts and registered rates where needed.
For BOTS, we start with model-token cost and add priced external enrichment calls recorded during the run.
We charge Brave Search at \$0.005 per request, WhoisXMLAPI WHOIS History preview at \$0.0258 per request, and WhoisXMLAPI WHOIS History purchase at \$1.29 per request.
We treat VirusTotal public API, DNS, and live WHOIS/RDAP as zero marginal cost.
Reported dollars are therefore model plus priced-enrichment costs only; they exclude Kubernetes, Splunk infrastructure, storage, and analyst review time.
We exclude costs for \texttt{bash}, \texttt{python}, and the three Splunk tool types.
We use a host-side cache for WhoisXMLAPI history responses to avoid repeated provider requests during evaluation; cached WhoisXMLAPI calls are still assigned their corresponding dollar value in reported tool-cost metrics so costs reflect expected tool use rather than cache state.

\section{Evaluation Results}
\label{sec:results}

We report results for GPT-5.5; GPT-5.6 Luna, Terra, and Sol; GPT-5.4 Mini; Claude Opus 4.8 and Fable 5; DeepSeek v4 Flash and Pro; Kimi K3; MiniMax M3; Qwen 3.7 Max and Qwen 3.5 Flash; GLM 5.2 and GLM 4.7 Flash; and Step 3.7 Flash.
The GPT-5.6 and Fable rows use high reasoning effort.
For BOTS v1 we additionally retain GPT-5.5 at high reasoning effort, the closest older comparison to Claude Opus 4.8, whose default setting is high effort.
For each GPT-5.6 variant we retain both the original operating point and a later run made after the evaluation account passed OpenAI Trusted Verification tier 1 around July 20, 2026. This operator-supplied account-state metadata is not encoded in the Inspect headers; the later logs were created on July 23 UTC. Tier 1 here is not the higher access tier that grants access to \texttt{gpt-5.5-cyber}.

\subsection{Offensive CTFs}
\label{sec:cybench-results}

Table~\ref{tab:cybench} summarizes Cybench hard-variant results for the shared model set.
GPT-5.5 remains the strongest operating point we observe, solving 94.1\% of challenges at \$1.16 per solved-equivalent challenge.
Among GPT-5.6 variants, Luna performs best at 79.5\%, followed by Terra at 65.8\%; Sol reaches 9.4\% after refusing 90.6\% of sample-epochs.
Fable refuses all 117 sample-epochs before any tool call, so its 0\% result is a policy-filter outcome rather than evidence of zero underlying capability.
DeepSeek v4 Flash provides the clearest illustration of cost scaling: replaying the same trace under a retrospective \$0.80 cap yields 76.1\% success, which rises to 86.4\% when the full \$2.10 budget is allowed.

\begin{table}[t]
\centering
\caption{Cybench operating points. GPT-5.6 and Fable use \texttt{high} reasoning effort. Refusals are counted as failures. GPT-5.5 had the highest outright success rate while DeepSeke v4 Flash was the most cost-efficient. Notably, frontier models like Claude Fable 5 and GPT-5.6-Sol had strong guardrails preventing them from completing the tasks. But even smaller GPT-5.6 models (Terra and Luna) had weaker guardrails enabling them to stay competitive despite refusals.}
\label{tab:cybench}
\resizebox{\linewidth}{!}{%
\begin{tabular}{llrrrrrr}
\toprule
Model & cost cap & success rate & solved equiv. & refusals & run cost & \$/solve & mean tool calls \\
\midrule
GPT-5.5 & \$2.10 & \textbf{94.1\%} & \textbf{36.7} & 7/117 & \$42.47 & \$1.16 & \textbf{20.6} \\
DeepSeek v4 Flash & \$2.10 & 86.4\% & 33.7 & 0/117 & \$48.88 & \$1.45 & 144.7 \\
GPT-5.6 Luna & \$2.10 & 79.5\% & 31.0 & 10/117 & \$40.66 & \$1.31 & 34.4 \\
Claude Opus 4.8 & \$2.10 & 76.2\% & 29.7 & 5/117 & \$94.58 & \$3.18 & 23.8 \\
DeepSeek v4 Flash & \$0.80 & 76.1\% & 29.7 & 0/117 & \$30.43 & \textbf{\$1.03} & 95.9 \\
GPT-5.6 Terra & \$2.10 & 65.8\% & 25.7 & 39/117 & \$36.63 & \$1.43 & 15.7 \\
DeepSeek v4 Pro & \$0.75 & 43.9\% & 17.1 & 0/114 & \$64.58 & \$3.78 & 21.3 \\
GPT-5.6 Sol & \$2.10 & 9.4\% & 3.7 & 106/117 & \$1.55 & \$0.42 & 1.4 \\
Claude Fable 5 & \$2.10 & 0.0\% & 0.0 & 117/117 & \$1.64 & -- & 0.0 \\
\bottomrule
\end{tabular}}
\end{table}

\subsection{SOC investigations}
\label{sec:bots-results}

Table~\ref{tab:bots} reports BOTS v1 full-agent results for the shared model set, with priced external-tool spend added to model-token spend.
Claude Opus 4.8 retains the highest defensive score at 9,666.7 of 10,300 points (93.9\%) and the best cost efficiency at \$2.98 per 1,000 points.
GPT-5.6 Terra and Sol follow at 92.1\% and 91.4\%; Fable reaches 88.4\% with the fewest non-submit tool calls, and Luna reaches 83.7\%.
All four exceed the standard GPT-5.5 row, but the public-benchmark contamination caveat below still applies.
DeepSeek v4 Flash supplies the clearest weak budget effect: doubling its cap from \$2.10 to \$4.20 moves its score from 73.0\% to only 73.9\%, while tool calls climb from 4,450 to 4,938.

\begin{table}[t]
\centering
\caption{BOTS v1 full-run results, including the latest high-effort GPT-5.6 and Fable runs. BOTS points use \texttt{bots\_points}; model + tools (\$) adds priced Brave Search and WhoisXMLAPI calls. "Refusal" columns measure how many sample-epochs out of 93 have been affected by guardrail refusals. Notably, Claude Fable 5 refused to complete a cybersecurity investigate or defensive tasks 5 times out of the 93 sample-epochs. }
\label{tab:bots}
\resizebox{\linewidth}{!}{%
\begin{tabular}{llrrrrrrr}
\toprule
Model & cost cap & BOTS points & score (\%) & binary acc. & refusals & model + tools (\$) & \$/1k pts & total tool calls \\
\midrule
Claude Opus 4.8 & \$2.10 & \textbf{9{,}666.7 / 10{,}300} & \textbf{93.9\%} & 96.8\% & 2/93 & \$28.80 & \textbf{\$2.98} & 603 \\
GPT-5.6 Terra & \$2.10 & 9{,}485.0 / 10{,}300 & 92.1\% & 95.7\% & 0/93 & \$33.66 & \$3.55 & 1{,}567 \\
GPT-5.6 Sol & \$2.10 & 9{,}416.7 / 10{,}300 & 91.4\% & \textbf{97.8\%} & 0/93 & \$44.29 & \$4.70 & 1{,}357 \\
Claude Fable 5 & \$2.10 & 9{,}108.3 / 10{,}300 & 88.4\% & 95.7\% & 5/93 & \$32.04 & \$3.52 & \textbf{309} \\
GPT-5.6 Luna & \$2.10 & 8{,}625.0 / 10{,}300 & 83.7\% & 93.5\% & 0/93 & \$29.66 & \$3.44 & 1{,}867 \\
GPT-5.5 (high effort) & \$2.10 & 8{,}383.3 / 10{,}300 & 81.4\% & 91.4\% & 0/93 & \$64.57 & \$7.70 & 1{,}580 \\
GPT-5.5 & \$2.10 & 8{,}345.0 / 10{,}300 & 81.0\% & 90.3\% & 0/93 & \$56.42 & \$6.76 & 1{,}481 \\
DeepSeek v4 Pro & \$2.10 & 8{,}013.3 / 10{,}300 & 77.8\% & 84.9\% & 0/93 & \$71.82 & \$8.96 & 3{,}363 \\
DeepSeek v4 Flash & \$4.20 & 7{,}610.0 / 10{,}300 & 73.9\% & 81.7\% & 0/93 & \$39.29 & \$5.16 & 4{,}938 \\
DeepSeek v4 Flash & \$2.10 & 7{,}518.3 / 10{,}300 & 73.0\% & 82.8\% & 0/93 & \$43.50 & \$5.79 & 4{,}450 \\
\bottomrule
\end{tabular}}
\end{table}

\subsection{Expanded models and provider operating conditions}
\label{sec:expanded-results}

Table~\ref{tab:expanded-results} mixes the additional model families into the comparison and preserves both GPT-5.6 operating points.
The post-verification GPT-5.6 runs are observational operating points rather than a controlled causal intervention: account status, run date, provider backend behavior, and other temporal effects may all differ.
The change is not uniformly favorable across benchmarks. For example, GPT-5.6 Sol rises from 9.4\% to 87.2\% on Cybench while its BOTS v1 score falls from 91.4\% to 78.8\%.
Among the additional models, Kimi K3 reaches 86.3\% on Cybench and 87.1\% of BOTS points.
MiniMax M3 reaches 76.5\% on Cybench with six provider/API failures retained as failures in the denominator: four exhausted API-timeout retries, one API connection error, and one response JSON-decode error.
This treatment measures the deployed model-provider system rather than accuracy conditional on receiving a valid response.

\begin{table}[t]
\centering
\caption{Expanded cross-benchmark operating points. ``Pre'' and ``post'' denote GPT-5.6 runs before and after OpenAI Trusted Verification tier 1, respectively. Cybench API errors are counted as failures. BOTS v1 uses \texttt{bots\_points}; binary accuracy remains secondary. GPT-5.4 Mini's earlier Cybench run used a \$0.80 cap, while the other added Cybench rows used \$2.10.}
\label{tab:expanded-results}
\small
\begin{tabular*}{\linewidth}{@{\extracolsep{\fill}}llrrrr@{}}
\toprule
Model & condition & Cybench & API errors & BOTS points & binary acc. \\
\midrule
GPT-5.6 Sol & pre & 9.4\% & 0/117 & 91.4\% & 97.8\% \\
GPT-5.6 Sol & post & 87.2\% & 0/117 & 78.8\% & 92.5\% \\
GPT-5.6 Terra & pre & 65.8\% & 0/117 & 92.1\% & 95.7\% \\
GPT-5.6 Terra & post & 85.5\% & 0/117 & 77.9\% & 88.2\% \\
GPT-5.6 Luna & pre & 79.5\% & 0/117 & 83.7\% & 93.5\% \\
GPT-5.6 Luna & post & 72.6\% & 0/117 & 71.2\% & 86.0\% \\
\midrule
Kimi K3 & July cohort & 86.3\% & 0/117 & 87.1\% & 94.6\% \\
MiniMax M3 & July cohort & 76.5\% & 6/117 & 67.2\% & 78.5\% \\
GLM 5.2 & July cohort & 73.5\% & 0/117 & 68.2\% & 80.6\% \\
Qwen 3.7 Max & July cohort & 71.8\% & 0/117 & 73.0\% & 86.0\% \\
Step 3.7 Flash & July cohort & 40.2\% & 0/117 & 62.1\% & 78.5\% \\
Qwen 3.5 Flash & July cohort & 35.9\% & 0/117 & 41.6\% & 51.6\% \\
GPT-5.4 Mini & mixed dates & 29.1\% & 0/117 & 75.2\% & 86.0\% \\
GLM 4.7 Flash & July cohort & 25.6\% & 0/117 & 38.4\% & 49.5\% \\
\bottomrule
\end{tabular*}
\end{table}

\begin{figure}[t]
\centering
\includegraphics[width=\linewidth]{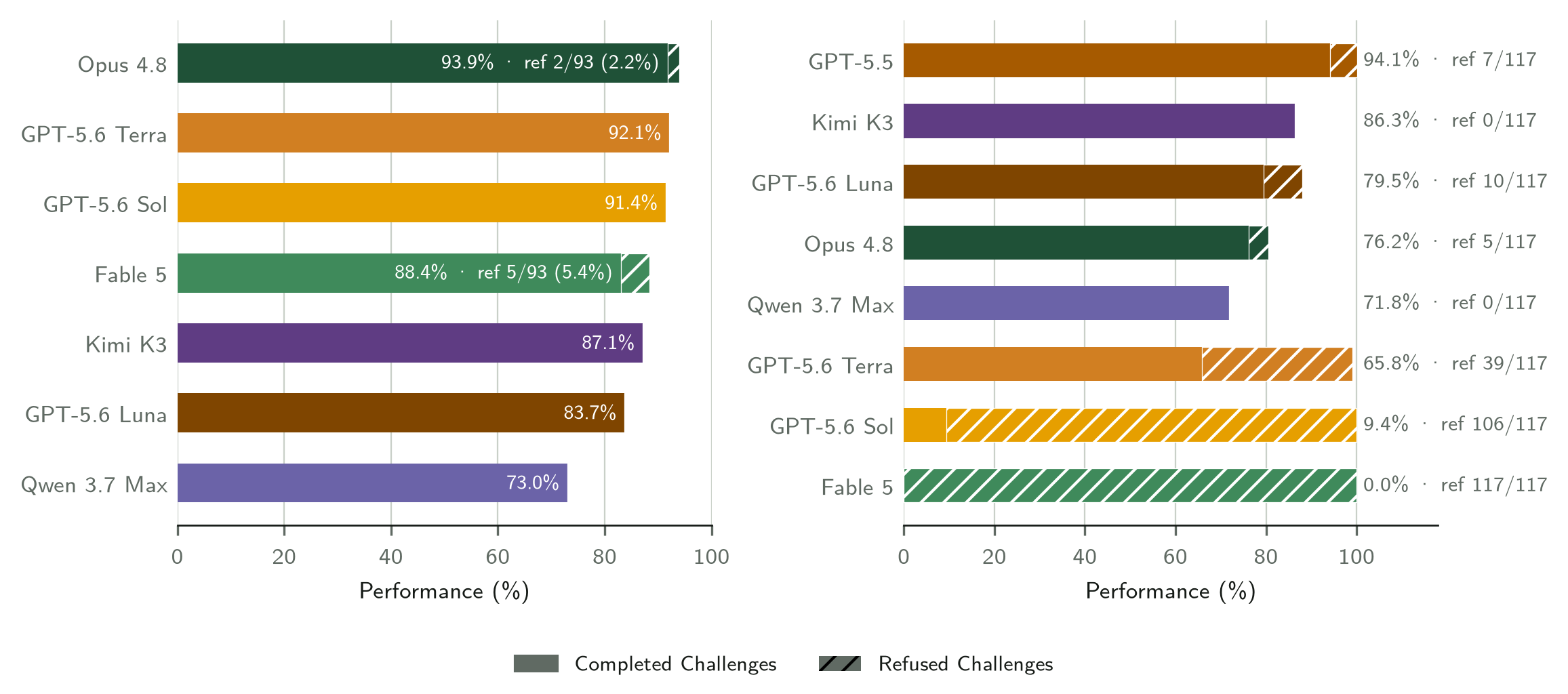}
\caption{Performance and policy refusals for selected current models, including Kimi K3 and Qwen 3.7 Max. Hatched segments show refusal events in both panels. \textbf{Left:} BOTS v1 refusal rates overlay the performance bars because affected sample-epochs may still earn points when a later attempt succeeds. \textbf{Right:} Cybench refusal outcomes are failures and therefore stack beyond, rather than overlap, solved share.}
\label{fig:gpt56-refusals}
\end{figure}

Figure~\ref{fig:gpt56-refusals} shows why performance and refusal context should be read together.
BOTS v1 refusal counts remain small and do not alter earned points; Fable, for example, earns 88.4\% despite refusal events in 5 of 93 sample-epochs.
On Cybench, policy gating instead dominates the lower GPT-5.6 and Fable results: Terra has 39 failed refusal outcomes, Sol 106, and Fable all 117.

\FloatBarrier

These full-agent scores should be interpreted alongside the contamination controls in Section~\ref{sec:bots-contamination}, not as a clean public leaderboard.

\section{Decontamination Controls}
\label{sec:bots-contamination}

\subsection{BOTS v1}
BOTS v1 is valuable because it supplies realistic Splunk data and official scoring, but it is also public and old.
We therefore run single-epoch no-tools probes that provide no Splunk, web, bash, Python, or other tools and ask the model to answer from the question text, optional official prerequisite Q\&A, and prior knowledge only.

The expanded results are strongly model-dependent.
Without prerequisite context, GPT-5.6 Sol scores 5,200/10,300 points (50.5\%), GPT-5.5 scores 54.9\%, and Claude Opus 4.8 scores 50.0\%, all with zero non-submit tool events.
By contrast, GPT-5.6 Luna, Terra, and Claude Fable 5 score 13.1\%, 18.4\%, and 14.6\%, respectively.
Official prerequisite Q\&A raises every no-tools score, most sharply for Sol (50.5\% to 77.2\%) and Claude Opus 4.8 (50.0\% to 74.8\%).
This spread argues against treating the probe as a uniform task-easiness correction: prompt-only answer recovery varies substantially by model and context, consistent with benchmark contamination but not diagnostic of its mechanism.

\begin{figure}[t]
\centering
\includegraphics[width=\linewidth]{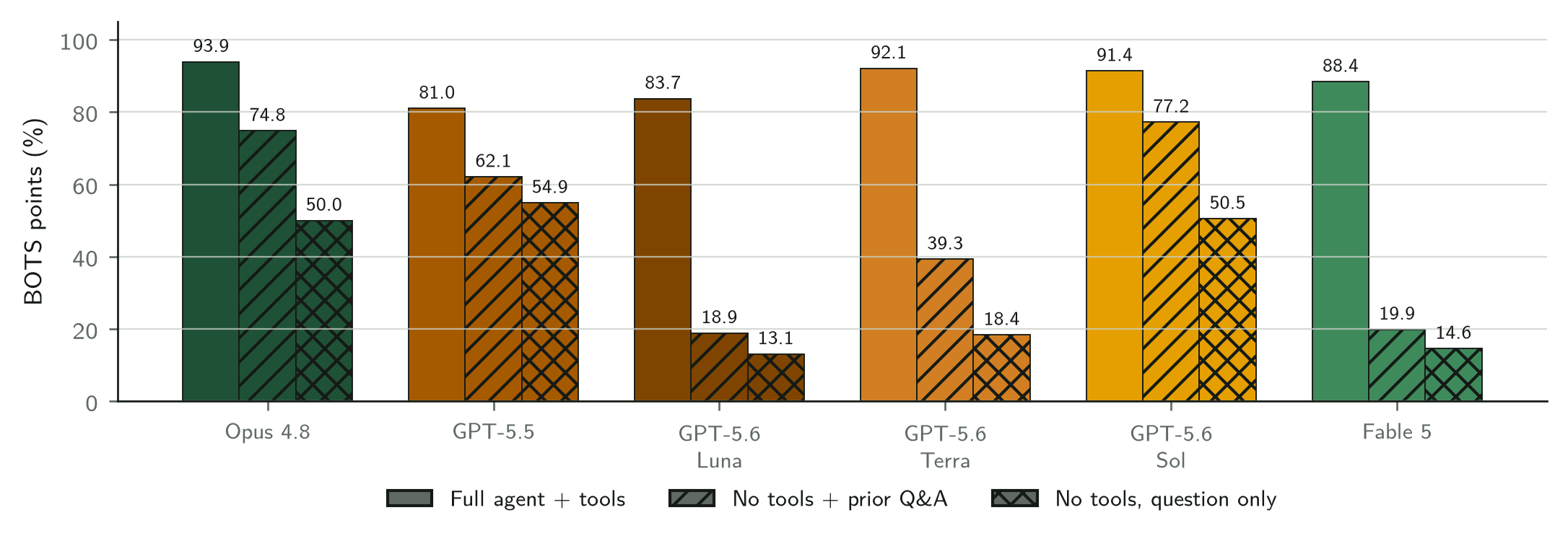}
\caption{BOTS v1 no-tools contamination probe. No-tools rows are single-epoch controls with zero non-submit tool events; full-agent baselines are three-epoch mean runs. Prior Q\&A denotes official prerequisite question-answer context.}
\label{fig:bots-decontamination}
\end{figure}

\begin{table}[t]
\centering
\caption{BOTS v1 decontamination-control results, scored with \texttt{bots\_points}.}
\label{tab:bots-no-tools}
\small
\begin{tabular*}{\linewidth}{@{\extracolsep{\fill}}lrrr@{}}
\toprule
Model & Full agent & No tools + prior Q\&A & Question only \\
\midrule
Claude Opus 4.8 & 93.9\% & 74.8\% & 50.0\% \\
GPT-5.5 & 81.0\% & 62.1\% & 54.9\% \\
GPT-5.6 Luna & 83.7\% & 18.9\% & 13.1\% \\
GPT-5.6 Terra & 92.1\% & 39.3\% & 18.4\% \\
GPT-5.6 Sol & 91.4\% & 77.2\% & 50.5\% \\
Claude Fable 5 & 88.4\% & 19.9\% & 14.6\% \\
\bottomrule
\end{tabular*}
\end{table}

These controls do not make BOTS useless; they make decontamination controls mandatory.
No-tools, perturbed-question, answer-before-query, private-holdout, or fresh-incident checks should accompany public SOC benchmarks before absolute scores are interpreted as live investigation skill.
BOTS remains useful as a reproducible harness and cost-accounting testbed, not as a clean leaderboard by itself.

\section{Scaling Findings}
\label{sec:scaling-findings}

\begin{figure}[t]
\centering
\widepaperfig{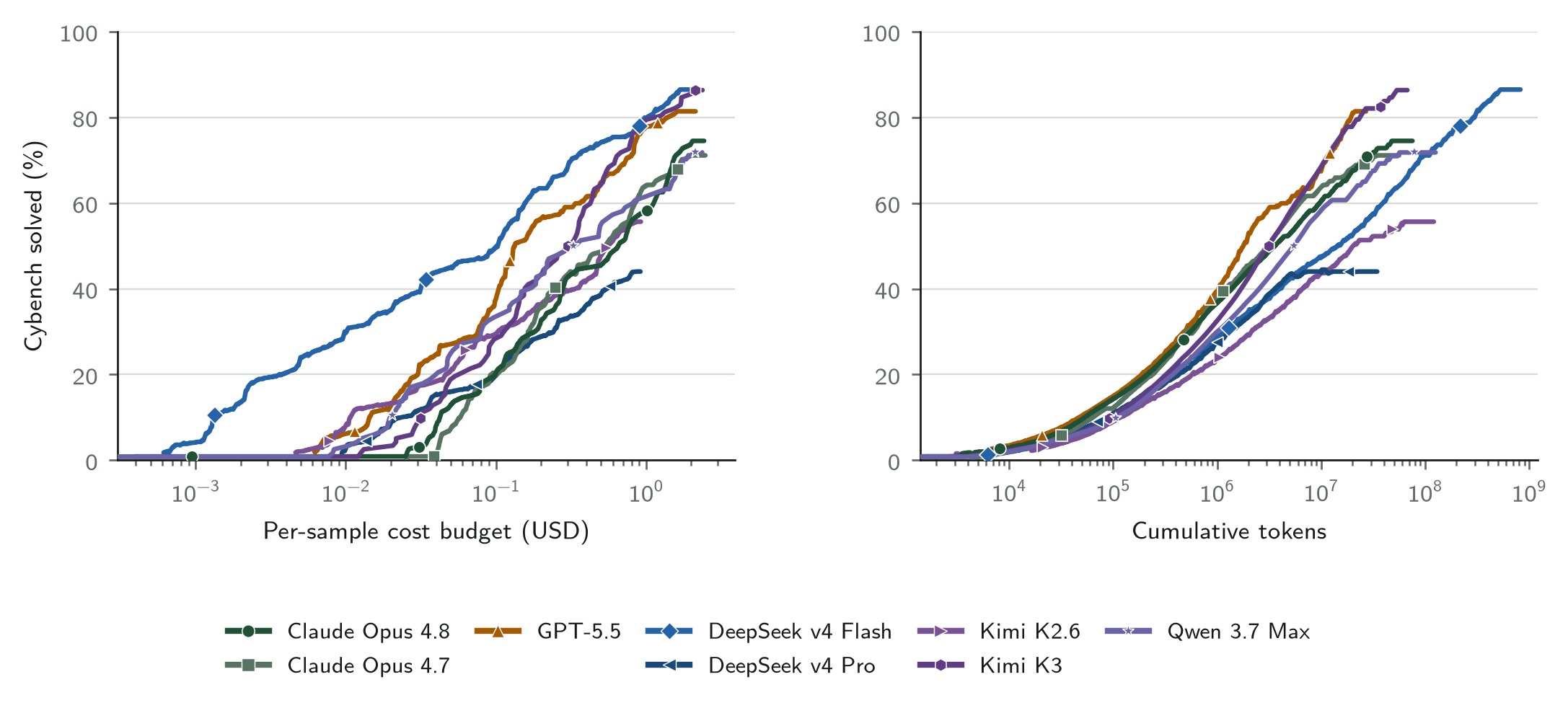}
\caption{Cybench scaling curves for the previously audited subset plus the July Kimi K3 and Qwen 3.7 Max traces.
\textbf{Left:} success versus retrospective per-sample cost budget (USD, log scale). \textbf{Right:} success versus cumulative tokens (input, output, cached, and reasoning). Success keeps rising toward the run limits: extra compute buys additional solves, most visibly for DeepSeek v4 Flash and Claude Opus 4.8, while GPT-5.5 solves most challenges at low per-sample spend.}
\label{fig:cybench-scaling-panels}
\end{figure}

\begin{figure}[t]
\centering
\widepaperfig{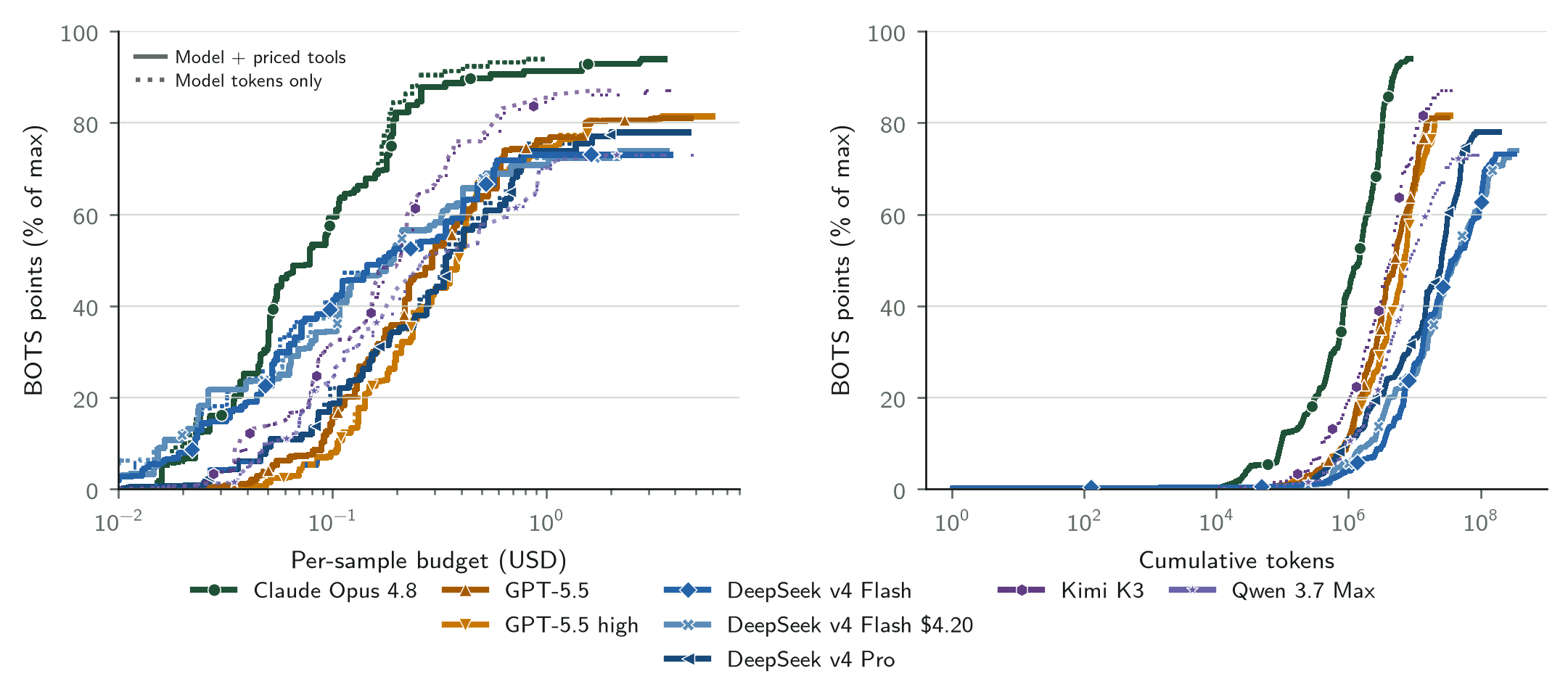}
\caption{BOTS v1 scaling curves for the previously audited full-agent subset plus Kimi K3 and Qwen 3.7 Max. Left: official BOTS points (\% of 10,300) versus a retrospective per-sample budget covering model-token plus priced-tool spend; dotted companions show the model-token-only budget, while solid curves include priced Brave Search and WhoisXMLAPI calls. Right: points versus cumulative tokens. Unlike Cybench, higher spend does not identify the best agent: Claude Opus 4.8 scores highest with roughly an order of magnitude fewer tokens than the DeepSeek runs, even though priced tools contribute materially to several runs' total cost.}
\label{fig:botsv1-scaling-panels}
\end{figure}

\begin{figure}[t]
\centering
\widepaperfig{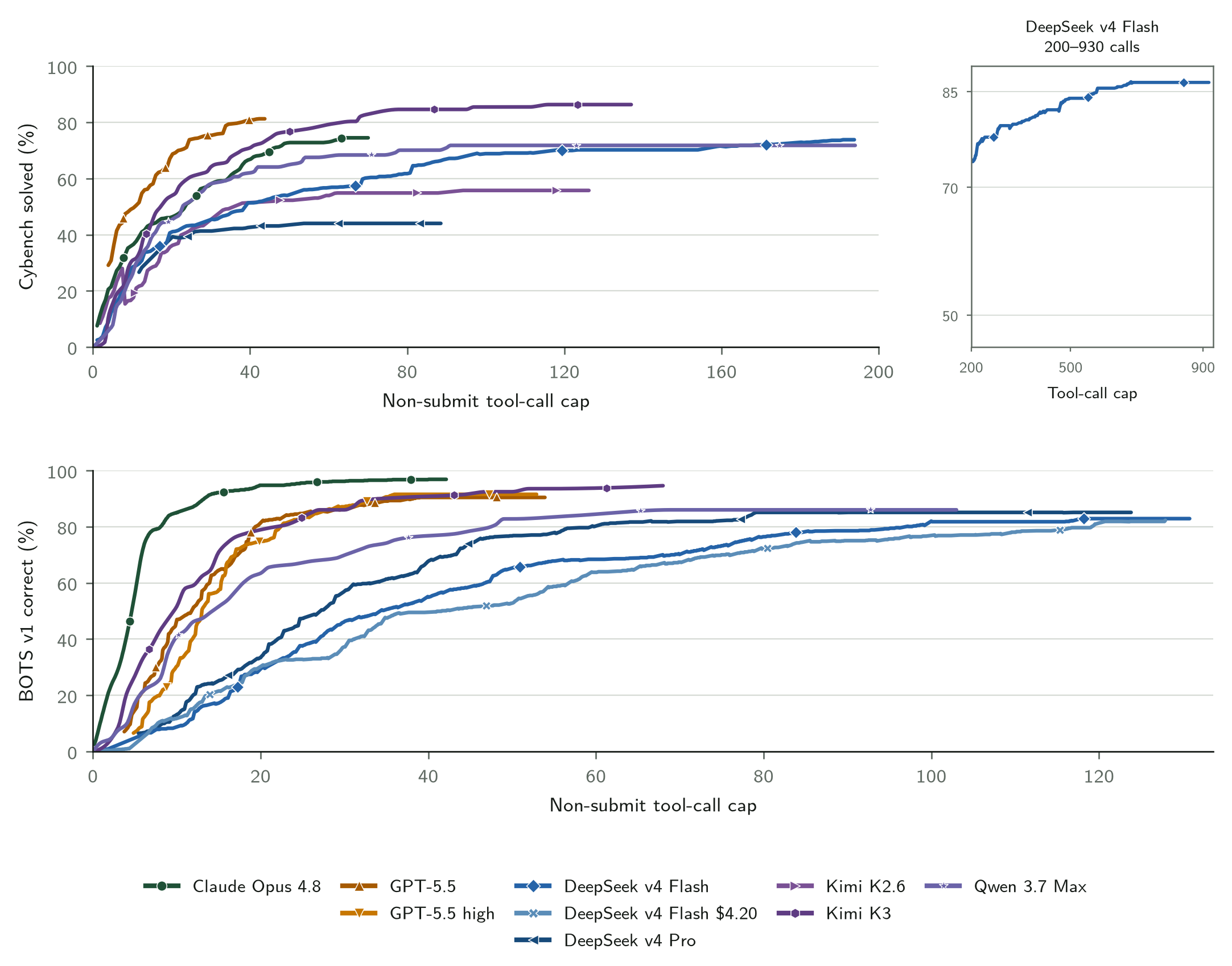}
\caption{Tool-call scaling: success at cap $x$ counts a sample-epoch only if it succeeded within $x$ non-submit tool calls. \textbf{Top:} Cybench success on a linear 0--200-call axis, with the inset continuing the DeepSeek v4 Flash trajectory over the 200--930-call long tail; most models saturate within ${\sim}50$ calls, but DeepSeek v4 Flash keeps extracting solves past 800 --- on offense, interaction volume can substitute for per-call quality. \textbf{Bottom:} BOTS v1 binary accuracy; the ordering inverts, with Claude Opus 4.8 highest at the fewest calls and the DeepSeek runs plateauing lower despite invoking more tools. High tool volume is not a proxy for defensive score.}
\label{fig:tool-call-scaling-panels}
\end{figure}

Budget-scaling claims are most credible when the comparison is paired: the same model, the same benchmark, and the same completed trace, with a lower budget imposed retrospectively.
We construct such comparisons by replaying each completed trace under a \$0.80 per-sample cap.
For Cybench, the cap follows model-call ledger events and marks a sample-epoch as failed once cumulative spend crosses \$0.80.
For BOTS v1, the cap applies to model-token plus priced-tool spend and zeroes a sample-epoch's points once total sample cost exceeds \$0.80.
These are descriptive replay analyses rather than new benchmark runs, but they avoid the confound of comparing unrelated models or unrelated traces.
Figures~\ref{fig:cybench-scaling-panels} and~\ref{fig:botsv1-scaling-panels} show the cost and token resource curves.
The Cybench panels also retain two additional previously audited runs, Claude Opus 4.7 and Kimi K2.6, and add the July Kimi K3 and Qwen 3.7 Max traces.
Cybench success rises with both per-sample dollar budget and cumulative token spend.
BOTS v1 has a different scaling shape: Claude Opus 4.8 reaches the highest score at lower cost and token volume, while higher-spend runs do not necessarily win.
Figure~\ref{fig:tool-call-scaling-panels} separately compares tool-call scaling across the two benchmarks.

\begin{table}[t]
\centering
\caption{Retrospective budget headroom above \$0.80 per sample. Deltas are full-trace score minus the same trace capped at \$0.80; bracketed values are 95\% paired bootstrap intervals over task IDs, in percentage points.}
\label{tab:scaling-headroom}
\resizebox{\linewidth}{!}{%
\begin{tabular}{lrrrr}
\toprule
Model & Cybench \$0.80$\rightarrow$full & Cybench $\Delta$ pp & BOTS v1 \$0.80$\rightarrow$full & BOTS v1 $\Delta$ pp \\
\midrule
GPT-5.5 & 91.5\%$\rightarrow$94.0\% & +2.6 [0.0, 6.0] & 74.1\%$\rightarrow$81.0\% & +6.9 [1.6, 13.6] \\
Claude Opus 4.8 (June run) & 55.6\%$\rightarrow$74.4\% & +18.8 [10.3, 29.1] & 91.3\%$\rightarrow$93.9\% & +2.6 [0.0, 7.1] \\
DeepSeek v4 Flash & 76.1\%$\rightarrow$86.3\% & +10.3 [4.3, 17.1] & 73.0\%$\rightarrow$73.0\% & +0.0 [0.0, 0.0] \\
DeepSeek v4 Pro & 43.9\%$\rightarrow$43.9\% & +0.0 [0.0, 0.0] & 73.8\%$\rightarrow$77.8\% & +4.0 [0.0, 9.6] \\
\bottomrule
\end{tabular}}
\end{table}

On Cybench, budget sensitivity is real but far from uniform.
Claude Opus 4.8 gains 18.8 percentage points when allowed to run past the \$0.80 cap, and DeepSeek v4 Flash gains 10.3 points; both bootstrap intervals are clearly positive.
GPT-5.5 gains only 2.6 points, because most of the challenges it solves already complete below \$0.80.
The near-zero Cybench delta for DeepSeek v4 Pro should not be read as evidence that the model cannot scale: the run was capped at a low prospective limit, so its observed trace offers little high-budget headroom to replay.

BOTS v1 shows a different shape.
Claude Opus 4.8 captures most of its points below \$0.80, and does so with only 603 non-submit tool calls over the full run.
GPT-5.5 retains some retrospective headroom above \$0.80, yet still trails Claude Opus 4.8 while spending almost twice as much and making more than twice as many tool calls; the high-effort GPT-5.5 row does not materially change this picture.
DeepSeek v4 Flash offers the cleanest direct budget check, since its two rows are prospective runs at different caps: the \$4.20-cap row exceeds the \$2.10-cap row by only 0.9 percentage points, with a paired bootstrap interval of $-4.9$ to $+6.9$ percentage points.
In this run, additional budget and tool volume did not buy a meaningful defensive improvement.

The practical lesson is not that spend never helps SOC agents; it is that the marginal dollar buys different things in the two workloads.
On Cybench, extra budget typically buys longer sandbox exploration, more command execution, and more chances to recover from false starts.
On BOTS v1, it often buys additional Splunk or enrichment calls after the agent has already missed the key field, prerequisite, or validation step.
Defensive evaluation should therefore scale not only dollars or tokens but also evidence quality: whether the agent finds the right telemetry, validates its answer, and pays for enrichment only when it changes the investigation.

\section{Limitations}
\label{sec:limitations}

Our experiments are observational rather than a prospectively randomized run matrix.
We analyze both prospectively fixed cost caps and retrospective caps applied to completed traces.
We therefore emphasize fixed-budget operating points over universal model rankings.
The expanded July comparison adds another non-invariant: OpenAI account verification status changed between the two GPT-5.6 run sets, while the Inspect logs do not encode that status. We report both operating points and avoid attributing their differences solely to verification. MiniMax M3 also exposes a system-level reliability effect: six provider/API failures are retained as Cybench failures rather than omitted or converted into conditional accuracy.
BOTS v1 rows are more uniform at the task level, though provider defaults and reasoning settings still vary across models.
BOTS v1 runs used a 250-message limit; this limit was hit by 8/93 DeepSeek v4 Flash sample-epochs at the \$2.10 cap, 14/93 at the \$4.20 cap, and 5/93 DeepSeek v4 Pro sample-epochs, so truncation may depress high-volume models.
The bootstrap intervals in Appendix~\ref{app:uncertainty} are descriptive robustness checks, not formal population-level inferences.

Our defensive analysis covers only the 31 scored BOTS v1 Po1s0n1vy/Cerber questions, after excluding the official warm-up, and uses a public Splunk dataset.
BOTS v2 and v3 extend v1: v2 covers a 51-question Frothly enterprise intrusion with a much larger endpoint, network, email, and web dataset, while v3 uses a 58-question cloud-heavy Frothly scenario spanning AWS CloudTrail, osquery, Windows, and network telemetry.
Comparable multi-model results for these later versions are not yet available, so extending the same cost-aware analysis to BOTS v2 and v3 is in scope for future work.
We therefore present the SOC conclusion as an evaluation-design result rather than a final claim about production SOC readiness.

\section{Artifact and Ethical Considerations}
\label{sec:ethics}

We use public cybersecurity benchmarks and sandboxed environments.
We report aggregate metrics and methodology only.
We do not publish secrets, API keys, or operational exploit instructions.

\section{Conclusion}
\label{sec:conclusion}

Security-agent evaluation should be cost-aware and workflow-specific.
Offensive CTFs and defensive SOC investigations stress different behaviors and show different scaling patterns.
We find that more test-time compute can improve CTF success for some frontier/shared models, especially Claude Opus 4.8 and DeepSeek v4 Flash under retrospective caps; the new GPT-5.6 and Fable runs also show that policy refusals can dominate offensive scores.
We also find that SOC work does not scale with raw budget in the same way; GPT-5.6 Terra and Sol approach the leading Claude Opus 4.8 score, while success still depends on disciplined tool use, telemetry navigation, and selective enrichment.
The expanded comparison shows that model names alone are insufficient operating-point identifiers: Kimi K3 is strong on both benchmarks, GPT-5.6 results shift across the two account-state periods, and MiniMax M3's provider/API failures measurably reduce deployed-system performance.
Public SOC benchmarks remain valuable, but the high and model-dependent no-tools BOTS v1 scores show that absolute scores require decontamination checks before they are used as capability claims.

\bibliography{references}

\appendix

\section{Evaluation Run Dates}
\label{app:evaluation-run-dates}

Tables~\ref{tab:cybench-eval-dates} and~\ref{tab:bots-eval-dates} report the UTC creation timestamp recorded in each source Inspect log used by the paper tables.
Rows that appear in multiple result, scaling, or robustness tables are listed once.
Where a row aggregates retry logs into the reported cost, the retry timestamps are included.
The retrospective cap rows are derived from the listed source trace and were not separate evaluation runs.

\begin{table}[htbp]
\centering
\caption{Cybench evaluation dates for tabled rows. Source timestamps are the \texttt{created} headers in the source Inspect logs used for the paper tables.}
\label{tab:cybench-eval-dates}
\small
\begin{tabular}{@{}>{\raggedright\arraybackslash}p{0.30\linewidth}>{\raggedright\arraybackslash}p{0.24\linewidth}>{\raggedright\arraybackslash}p{0.38\linewidth}@{}}
\toprule
Table row & Source evaluation timestamp (UTC) & Notes \\
\midrule
GPT-5.5, \$2.10 & 2026-05-14 18:34:57; 20:45:04; 20:52:17; 21:03:03 & Three quota-failed retries plus final success; used in Tables~\ref{tab:cybench}, \ref{tab:scaling-headroom}, and \ref{tab:cybench-bootstrap} \\
GPT-5.6 Luna, \$2.10 & 2026-07-09 20:44:03 & Successful retry; cost reconstructed from cumulative tokens \\
GPT-5.6 Terra, \$2.10 & 2026-07-10 06:43:26 & High reasoning effort \\
GPT-5.6 Sol, \$2.10 & 2026-07-10 15:35:10 & High reasoning effort; 90.6\% refused \\
Claude Fable 5, \$2.10 & 2026-07-13 18:04:21 & High reasoning effort; all sample-epochs content-filtered \\
DeepSeek v4 Flash, \$2.10 & 2026-05-19 16:35:07 & Same trace also supplies the retrospective \$0.80 row \\
Claude Opus 4.8, \$2.10 & 2026-07-13 18:17:02 & Latest main-table and refusal-figure run \\
Claude Opus 4.8, \$2.10 (June run) & 2026-06-22 18:37:21; 2026-06-23 19:08:17 & Scaling and bootstrap analyses \\
DeepSeek v4 Pro, \$0.75 & 2026-05-15 19:38:26; 23:35:01; 23:44:23 & Same-day error retries plus final success \\
GPT-5.6 Luna/Terra/Sol, post-verification & 2026-07-23 20:38:55; 20:38:56; 20:38:56 & Table~\ref{tab:expanded-results}; Trusted Verification status is operator-supplied \\
Kimi K3; Qwen 3.7 Max; MiniMax M3; Step 3.7 Flash & 2026-07-23 20:38:53; 20:38:53; 20:38:54; 20:38:54 & Table~\ref{tab:expanded-results}; MiniMax retains six API errors as failures \\
GLM 5.2; Qwen 3.5 Flash; GLM 4.7 Flash & 2026-07-23 20:38:52; 20:38:55; 20:38:55 & Table~\ref{tab:expanded-results} \\
GPT-5.4 Mini, \$0.80 & 2026-05-15 06:49:32 & Table~\ref{tab:expanded-results}; earlier Cybench run \\
\bottomrule
\end{tabular}
\end{table}

\begin{table}[htbp]
\centering
\caption{BOTS v1 evaluation dates for tabled rows. Source timestamps are the \texttt{created} headers in the source Inspect logs used for the paper tables.}
\label{tab:bots-eval-dates}
\small
\begin{tabular}{@{}>{\raggedright\arraybackslash}p{0.35\linewidth}>{\raggedright\arraybackslash}p{0.24\linewidth}>{\raggedright\arraybackslash}p{0.33\linewidth}@{}}
\toprule
Table row & Source evaluation timestamp (UTC) & Notes \\
\midrule
Claude Opus 4.8, full agent, \$2.10 & 2026-06-15 04:30:45 & Tables~\ref{tab:bots}, \ref{tab:scaling-headroom}, \ref{tab:bots-bootstrap}, and \ref{tab:bots-no-tools} \\
GPT-5.6 Terra, full agent, \$2.10 & 2026-07-13 19:54:58 & High reasoning effort; Table~\ref{tab:bots} \\
GPT-5.6 Sol, full agent, \$2.10 & 2026-07-13 19:55:19 & High reasoning effort; Table~\ref{tab:bots} \\
Claude Fable 5, full agent, \$2.10 & 2026-07-13 18:19:40 & High reasoning effort; Table~\ref{tab:bots} \\
GPT-5.6 Luna, full agent, \$2.10 & 2026-07-13 17:43:15 & High reasoning effort; Table~\ref{tab:bots} \\
GPT-5.5 high effort, full agent, \$2.10 & 2026-06-15 06:09:42 & Tables~\ref{tab:bots} and \ref{tab:bots-bootstrap} \\
GPT-5.5, full agent, \$2.10 & 2026-06-14 21:12:20 & Tables~\ref{tab:bots}, \ref{tab:scaling-headroom}, \ref{tab:bots-bootstrap}, and \ref{tab:bots-no-tools} \\
DeepSeek v4 Pro, full agent, \$2.10 & 2026-06-14 23:14:52 & Tables~\ref{tab:bots}, \ref{tab:scaling-headroom}, and \ref{tab:bots-bootstrap} \\
DeepSeek v4 Flash, full agent, \$4.20 & 2026-06-15 04:59:44 & Tables~\ref{tab:bots} and \ref{tab:bots-bootstrap} \\
DeepSeek v4 Flash, full agent, \$2.10 & 2026-06-14 23:07:00 & Tables~\ref{tab:bots}, \ref{tab:scaling-headroom}, and \ref{tab:bots-bootstrap} \\
Claude Opus 4.8, no tools, prereq context & 2026-06-23 21:32:08 & Table~\ref{tab:bots-no-tools} \\
Claude Opus 4.8, no tools, question only & 2026-06-23 21:51:42 & Table~\ref{tab:bots-no-tools} \\
GPT-5.5, no tools, prereq context & 2026-06-23 22:21:31 & Table~\ref{tab:bots-no-tools} \\
GPT-5.5, no tools, question only & 2026-06-23 22:43:04 & Table~\ref{tab:bots-no-tools} \\
GPT-5.6 Luna, no tools, prior Q\&A / question only & 2026-07-17 04:22:27; 05:55:04 & Table~\ref{tab:bots-no-tools} \\
GPT-5.6 Terra, no tools, prior Q\&A / question only & 2026-07-17 05:55:04; 06:24:56 & Table~\ref{tab:bots-no-tools} \\
GPT-5.6 Sol, no tools, prior Q\&A / question only & 2026-07-17 06:24:56; 06:54:08 & Table~\ref{tab:bots-no-tools} \\
Claude Fable 5, no tools, prior Q\&A / question only & 2026-07-17 06:54:07; 07:22:40 & Table~\ref{tab:bots-no-tools} \\
GPT-5.6 Luna/Terra/Sol, post-verification & 2026-07-23 08:20:08 & Table~\ref{tab:expanded-results}; three separate logs with the same creation second \\
Kimi K3; MiniMax M3; GLM 5.2; Qwen 3.7 Max & 2026-07-23 08:20:08 & Table~\ref{tab:expanded-results}; separate full-run logs \\
Qwen 3.5 Flash; GPT-5.4 Mini; GLM 4.7 Flash & 2026-07-23 08:20:08 & Table~\ref{tab:expanded-results}; separate full-run logs \\
Step 3.7 Flash retry chain & 2026-07-23 08:20:08; 2026-07-24 07:07:36; 15:18:04; 15:27:02 & Table~\ref{tab:expanded-results}; final retry completed 93/93 \\
\bottomrule
\end{tabular}
\end{table}
\FloatBarrier

\section{Uncertainty Method}
\label{app:uncertainty}

We report descriptive bootstrap intervals.
For Cybench, each replicate resamples challenge IDs with replacement and retains the three epochs for each selected challenge.
For BOTS v1, each replicate resamples question IDs with replacement and retains the three epochs and official point weights for each selected question.
For retrospective cap contrasts, each replicate compares the capped and uncapped values for the same sampled task IDs.
Paired contrasts resample only task IDs common to both compared runs.
The GPT-5.6, Fable, and latest July Opus rows are not included in these previously computed bootstrap or scaling analyses.

\section{Cybench Robustness}
\label{app:cybench-robustness}

\begin{table}[htbp]
\centering
\caption{Cybench challenge-level bootstrap intervals. The DeepSeek v4 Flash \$0.80 row uses the same retrospective cap as Table~\ref{tab:cybench}. Last-digit differences from the main table reflect independent recomputation and rounding from logs.}
\label{tab:cybench-bootstrap}
\small
\begin{tabular*}{\linewidth}{@{\extracolsep{\fill}}llrr@{}}
\toprule
Model & cost cap & success rate & 95\% bootstrap interval \\
\midrule
GPT-5.5 & \$2.10 & 94.0\% & 87.2--99.2\% \\
DeepSeek v4 Flash & \$2.10 & 86.3\% & 76.9--94.9\% \\
DeepSeek v4 Flash & \$0.80 & 76.1\% & 63.2--87.2\% \\
Claude Opus 4.8 (June run) & \$2.10 & 74.4\% & 61.5--86.3\% \\
DeepSeek v4 Pro & \$0.75 & 43.9\% & 29.8--57.9\% \\
\bottomrule
\end{tabular*}
\end{table}

The main scaling contrasts are reported in Table~\ref{tab:scaling-headroom}.
For model-ranking context, GPT-5.5 exceeds the higher-budget DeepSeek v4 Flash row by 7.7 percentage points, with interval +1.7 to +15.4 percentage points.
Claude Opus 4.8 does not close the gap to those leaders: paired contrasts are GPT-5.5 minus Claude Opus 4.8 = +19.7 percentage points [+9.4, +30.8] and higher-budget DeepSeek v4 Flash minus Claude Opus 4.8 = +12.0 percentage points [+1.7, +23.1].

\section{BOTS v1 Robustness}
\label{app:bots-robustness}

\begin{table}[htbp]
\centering
\caption{BOTS v1 question-level bootstrap intervals.}
\label{tab:bots-bootstrap}
\small
\begin{tabular*}{\linewidth}{@{\extracolsep{\fill}}llrr@{}}
\toprule
Model & cost cap & BOTS score (\%) & 95\% bootstrap interval \\
\midrule
Claude Opus 4.8 & \$2.10 & 93.9\% & 82.7--99.5\% \\
GPT-5.5 high effort & \$2.10 & 81.4\% & 67.4--94.1\% \\
GPT-5.5 & \$2.10 & 81.0\% & 64.1--96.4\% \\
DeepSeek v4 Pro & \$2.10 & 77.8\% & 61.0--93.1\% \\
DeepSeek v4 Flash & \$4.20 & 73.9\% & 53.2--91.9\% \\
DeepSeek v4 Flash & \$2.10 & 73.0\% & 52.7--91.1\% \\
\bottomrule
\end{tabular*}
\end{table}

Claude Opus 4.8's paired margin over the standard GPT-5.5 row is 12.8 percentage points, with a question-level paired bootstrap interval of +0.9 to +27.4 percentage points.
Its margin over GPT-5.5 high effort is 12.5 percentage points, with interval +2.4 to +23.7 percentage points.
The DeepSeek v4 Flash \$4.20-cap row exceeds the \$2.10-cap row by only 0.9 percentage points, with interval $-4.9$ to $+6.9$ percentage points, so that budget change is inconclusive.

\section{BOTS v1 Sequential-Context Check}
\label{app:bots-context}

We exclude the official BOTS v1 warm-up question and evaluate 31 scored questions.
Twenty-three of those questions have explicit dependencies on earlier case facts and receive prerequisite context in the prompt.
This matches the benchmark's sequential investigation structure, but we interpret it as follow-up investigation with known case state rather than cold-start reconstruction.
Table~\ref{tab:bots-dependencies} shows the direct prerequisite chart; the prompt preamble expands these edges transitively.

\begin{table}[!htbp]
\centering
\caption{BOTS v1 direct prerequisite chart from the benchmark metadata. Only dependent scored questions are shown; independent scored questions and the warm-up are omitted. Source: benchmark metadata.}
\label{tab:bots-dependencies}
\small
\begin{tabular*}{\linewidth}{@{\extracolsep{\fill}}llll@{}}
\toprule
Dependent Q & Direct prior Q(s) & Dependent Q & Direct prior Q(s) \\
\midrule
\multicolumn{4}{l}{\textit{Po1s0n1vy web defacement}} \\
Q104 & Q101 & Q112 & Q111 \\
Q105 & Q104 & Q113 & Q106 \\
Q106 & Q104 & Q114 & Q108 \\
Q107 & Q106 & Q115 & Q114 \\
Q108 & Q101 & Q116 & Q114 \\
Q109 & Q108 & Q117 & Q114 \\
Q110 & Q109 & Q118 & Q116 \\
Q111 & Q106 & Q119 & Q114 \\
\midrule
\multicolumn{4}{l}{\textit{Cerber ransomware}} \\
Q202 & Q200 & Q208 & Q204 \\
Q203 & Q202 & Q210 & Q203 \\
Q206 & Q200 & Q211 & Q210 \\
Q207 & Q206 & & \\
\bottomrule
\end{tabular*}
\end{table}
\FloatBarrier

For Claude Opus 4.8, independent questions score 1,445/1,500 points (96.3\%) and dependent questions score 8,221.7/8,800 points (93.4\%).
The standard GPT-5.5 row scores 1,345/1,500 points (89.7\%) on independent questions and 7,000/8,800 points (79.5\%) on dependent questions.
Claude's mean-reduced point loss decomposes into 133.3 points from hint penalties and 500 points from misses; GPT-5.5 loses 205 points to hint penalties and 1,750 points to misses.

\end{document}